\documentstyle[12pt]{article}
\input epsf   
\begin{document}
\title
{Higgs boson production under the resonance threshold at LEP II}
\author
{E.Boos, M.Dubinin, L.Dudko \\
{\small \it Institute for Nuclear Physics, Moscow State University} \\  
{\small \it 119899 Moscow, Russia} }
\date{} 
\maketitle

\begin{abstract}
We consider the possibility of Higgs boson detection at LEP II
under the resonance threshold ($\sqrt{s}<m_H+m_Z$) in the framework
of complete tree level approach to the calculation of the 
$e^+ e^- \rightarrow \nu \bar \nu b \bar b$ amplitude, simulating b-quark 
fragmentation to hadrons and taking into account typical 
detector properties. At the energy below the $2m_Z$ threshold $\sqrt{s}=$ 
175 GeV Higgs boson production under the $m_H+m_Z$ threshold is 
almost background free. 
\end{abstract}
\begin{picture}(1,1)
\put(300,340){INP MSU 96-2/409}  
\put(300,325){SNUTP 96-004}
\end{picture}
\section{Introduction}

The possibility of Higgs boson detection in the four
fermion final states at LEP II has been investigated recently
on the level of complete tree level calculations, when the full 
set of diagrams (signal and irreducible background) is considered. 
In the papers \cite{BSSS1,BSSS2,BSSS3} complete tree level calculation 
for the Higgs signal and irreducible backgrounds in the semileptonic 
four fermion processes
\begin{eqnarray}
&e^+e^- \rightarrow \mu^+\mu^- b \bar b& \\
&e^+e^- \rightarrow \nu \bar \nu b \bar b& \\
&e^+e^- \rightarrow e^+e^- b \bar b&
\end{eqnarray}
has been done. In the processes (2) and (3) there are two types of
signal diagrams for Higgs boson production (see Fig.1): 1) Higgs 
bremsstrahlung from the $Z$-boson line; 2) Higgs production by $WW$ or 
$ZZ$ fusion.

Higgs bremsstrahlung \cite{brems} and fusion \cite{fusion} mechanisms
taken separately (as the noninterfering amplitudes $e^+ e^- \rightarrow HZ$ 
and $e^+ e^- \rightarrow \nu_e \bar \nu_e H$) have been discussed 
for a long time. However it is much better to consider both mechanisms 
as the interfering parts of one amplitude (performing coherent summation of 
corresponding Feynman diagrams). The interplay of two 
mechanisms is especially interesting
near the threshold energy $\sqrt{s}=m_H+m_Z$ in the $\nu_e \bar \nu_e b \bar 
b$ channel, when the contributions
of Higgsstrahlung and fusion diagrams are of the same order and the
interference term is positive and not negligible. 

In particular it was shown \cite{BSSS2} that in the process (2) at the 
energies 
$\sqrt{s} < m_H+m_Z$ (under the threshold $m_H+m_Z$) fusion mechanism is more
important in the channel $e^+ e^- \rightarrow \nu_e \bar \nu_e b \bar b$ 
than Higgsstrahlung and can lead to observable events at
LEP II luminosities. The number of events decreases as we go down
in energy from the point $\sqrt{s}=m_H+m_Z$ (or, equivalently, go
up in Higgs mass from the point $m_H=\sqrt{s}-m_Z$), but nevertheless
in the mass interval of about 10 GeV under the threshold it could be
possible to observe from four to ten Higgs production events/year.
In other words, LEP II gives the possibility to look for the Higgs
boson with the mass $m_H=\sqrt{s}-m_Z+\Delta m$, where $\Delta m$ is
about 10 GeV. 
Higgs peak could be observed in the invariant mass distribution of two
b-jets and for this reason direct experimental reconstruction of two b-jets 
is very critical for signal separation \cite{BD}.

The importance of
fusion and interference terms in the threshold region has been
mentioned in \cite{BD} and investigated in more details in \cite{BSSS2}
by means of Monte-Carlo simulation in the framework of complete tree 
level approach (23 diagrams in the $\nu_e \bar \nu_e b \bar b$ channel, 11 
diagrams in the $\nu_{\mu} \bar \nu_{\mu} b \bar b$ and $\nu_{\tau} \bar 
\nu_{\tau} b \bar b$ channels). Semianalytic results for 
the total cross section 
and analytic distributions for two signal diagrams and interference 
between them can be found in \cite{KKZ} where the $2 \rightarrow 3$ 
body approximation $e^+ e^- \rightarrow \nu_e \bar \nu_e H$ has been used.
Complete tree level semianalytic results for the $2 \rightarrow 4$ body
process (1) (25 diagrams) were obtained in \cite{Bardin}. However they 
are not extended yet to the case of fusion mechanisms.

The main purpose of present paper is the investigation of signal-background
ratio in the process (2) under the threshold $m_H+m_Z$ in the framework of 
complete 
tree level approach. The number of signal events is small at 
LEP II 
luminosities and for this reason it is very important to have detailed
understanding of the background taking into account some realistic 
properties of detector environment. For this reason we simulate b-quark
fragmentation and employ some typical detector model for the calculation 
of $M(b \bar b)$ distribution.  

Higgs production by $ZZ$ fusion in the process (3) will also take
place. However, the irreducible background diagrams give the cross
section about 100 times larger than the signal and in this case we need
nontrivial and complicated procedure of signal separation \cite{BSSS3}.

\section{Cross sections}

At present time several approaches exist (with the algorithmic
realizations in the form of MC integrators or event generators) 
for the calculation of four fermion states at complete tree level. 
Description of general strategies can be found in \cite{general}.
We used CompHEP package \cite{CompHEP} for complete tree level
calculation of the signal and irreducible background. As usual,
all possible squared diagrams and interferences between them
(including signal-background interferences) were calculated. Fermion
masses were kept nonzero in the amplitude calculation and four
particle phase space generation. Two signal
diagrams and 21 bias graphs for the process $e^+ e^- \rightarrow
\nu_e \bar \nu_e b \bar b$ are shown in Fig.1. In the case of
muon and tau neutrino in the final state the complete set of
diagrams contains one signal and 10 bias graphs. All channels
($\nu_e, \nu_\mu, \nu_\tau$) were taken into account in our
simulation. Total cross sections of these processes at the
energies $\sqrt{s}=175$ GeV and $\sqrt{s}=205$ GeV are shown in 
Table 1 for Higgs boson masses $m_H=$ 85, 90, 95 GeV and $m_H=$
115, 120, 125 GeV respectively.

\begin{table}[h]
\begin{center}
\begin{tabular}{|c|c|c|c|c|c|c|c|c|c|}               \hline
\multicolumn{10}{|c|}{$\sqrt{s}=$175 GeV}\\           \hline
$m_H$,GeV&\multicolumn{3}{|c|}{85}&\multicolumn{3}{|c|}{90}&
\multicolumn{3}{|c|}{95}     \\ \hline 
$\sigma_{tot}$ [fb] & \multicolumn{3}{|c|} { } & \multicolumn{3}{|c|} { }& 
\multicolumn{3}{|c|} { } \\
 fixed $\Gamma$& \multicolumn{3}{|c|}{34.8} 
&\multicolumn{3}{|c|}{20.0} & \multicolumn{3}{|c|}{16.0}  \\ \hline 
$\sigma_{tot}$ [fb] & \multicolumn{3}{|c|} { } & \multicolumn{3}{|c|} { }& 
\multicolumn{3}{|c|} { } \\
 overall $\Gamma$   &\multicolumn{3}{|c|}{33.7} 
&\multicolumn{3}{|c|}{14.1} & \multicolumn{3}{|c|}{15.3}  \\ \hline 
channel &$\nu_e$&$\nu_\mu$&$\nu_\tau$&$\nu_e$&$\nu_\mu$&$\nu_\tau$&
$\nu_e$&$\nu_\mu$&$\nu_\tau$ \\ \hline
$\sigma_{tot}$ [fb] &&&&&&&&& \\
fixed $\Gamma$ & 17.9  &  8.4   &   8.4 & 11.2 & 4.4    & 4.4   & 
 8.7  & 3.6    & 3.6 \\ \hline
$\sigma_{tot}$ [fb] &&&&&&&&& \\
overall $\Gamma$ &17.3 &  8.2  &   8.2 & 6.9 & 3.6   & 3.6  & 
 8.2  & 3.6  & 3.6 \\ \hline
\multicolumn{10}{|c|}{$\sqrt{s}=$205 GeV} \\  \hline
$m_H$,GeV&\multicolumn{3}{|c|}{115}&\multicolumn{3}{|c|}{120}&
\multicolumn{3}{|c|}{125}     \\ \hline
$\sigma_{tot}$ [fb] & \multicolumn{3}{|c|} { } & \multicolumn{3}{|c|} { }& 
\multicolumn{3}{|c|} { } \\
 fixed $\Gamma$  & \multicolumn{3}{|c|}{98.6} 
&\multicolumn{3}{|c|}{92.0} & \multicolumn{3}{|c|}{89.9}  \\ \hline 
$\sigma_{tot}$ [fb] & \multicolumn{3}{|c|} { } & \multicolumn{3}{|c|} { }& 
\multicolumn{3}{|c|} { } \\
 overall $\Gamma$ &\multicolumn{3}{|c|}{98.2} 
&\multicolumn{3}{|c|}{91.6} & \multicolumn{3}{|c|}{89.6}  \\ \hline 
channel&$\nu_e$&$\nu_\mu$&$\nu_\tau$&$\nu_e$&$\nu_\mu$&$\nu_\tau$&
$\nu_e$&$\nu_\mu$&$\nu_\tau$ \\ \hline
$\sigma_{tot}$ [fb] &&&&&&&&& \\
fixed $\Gamma$ &39.1 & 29.8  & 29.8 & 35.7 & 28.2    & 28.2   &
 34.2  & 27.9    & 27.9 \\ \hline
$\sigma_{tot}$ [fb] &&&&&&&&& \\
overall $\Gamma$ & 38.8 & 29.7 &   29.7 & 35.4 & 28.1  
& 28.1  & 33.9 & 27.8 & 27.8 \\ \hline
\end{tabular}
\end{center}
\caption{Total cross sections for the  
the processes $e^+ e^- \rightarrow \nu \bar \nu b \bar b$, $\nu=\nu_e, 
\nu_{\mu}, \nu_{\tau}$ calculated using two prescriptions for insertion 
of exact propagator in the amplitude. Parameter 
values $m_b=4.3$ GeV, $m_Z=91.19$ GeV, $\Gamma_Z=$ 2.50 GeV, 
$sin^2 \vartheta_w=$ 0.225, $\alpha=1/128$ were used. } 
\end{table}

Two methods for the insertion of finite width in the $H$ and $Z,W$
propagators were used. The "fixed width" method implements the replacement
$$
\frac{1}{k^2-m^2+i\epsilon} \rightarrow \frac{1}{k^2-m^2+im\Gamma}
$$
in the resonant graphs only. Fixed width method violates  
the gauge invariance of the amplitude and at the same time does not affect 
nonresonant 
graphs. In the "overall" prescription the whole amptlitude is multiplied by 
the factor 
$$
\frac{k^2-m^2+i\epsilon}{k^2-m^2+im\Gamma}
$$
preserving gauge invariance but obviously underestimating the
contribution of nonresonant graphs. Besides this we would like to indicate 
one more drawback of the "overall" prescription. For instance, at the energy
$\sqrt{s}=$ 175 GeV and Higgs boson mass $m_H=$ 90 GeV (see Table 1) the
difference of results obtained by two methods is 25 \%. In
this case large contribution from the $Z$ resonance is suppressed
by the "overall" factor of Higgs propagator and vice versa. The case
when $Z$ and $H$ peaks are close to each other gives one more example
of the situation when the "overall" method cannot be applied for meaningful 
calculation.
General discussion of the problem how to use boson propagators in the
tree diagrams calculation at high energy can be found in \cite{width};
although consistent results can be obtained in many cases by using different 
methods, the problem seems far from being understood completely.

\begin{table}[h]
\begin{center}
\begin{tabular}{|c|c|c|c|c|}               \hline
\multicolumn{5}{|c|}{$\sqrt{s}=$175 GeV}\\           \hline
\multicolumn{2}{|c|}{$m_H$,GeV}&85&90&95     \\ \hline 
\multicolumn{2}{|c|}{ Fusion mechanism}&4.2&3.4&2.7     \\ \hline 
\multicolumn{2}{|c|}{ Higgsstrahlung mechanism}&5.5&1.5&0.7     \\ \hline 
\multicolumn{2}{|c|}{ Interference}&5.0&2.9&1.9     \\ \hline  
\multicolumn{2}{|c|}{ Total signal cross section}&14.5&7.8&5.3     \\ \hline 
\multicolumn{2}{|c|}{ Background cross section} & 
\multicolumn{3}{|c|}{3.15}    \\ \hline 
\multicolumn{5}{|c|}{$\sqrt{s}=$205 GeV}\\           \hline
\multicolumn{2}{|c|}{$m_H$,GeV}&115&120&125     \\ \hline 
\multicolumn{2}{|c|}{ Fusion mechanism}&3.2&2.6&2.0     \\ \hline 
\multicolumn{2}{|c|}{ Higgsstrahlung mechanism}&2.2&0.6&0.3     \\ \hline 
\multicolumn{2}{|c|}{ Interference}&2.8&1.6&1.1     \\ \hline  
\multicolumn{2}{|c|}{ Total signal cross section}&8.1&4.8&3.4     \\ \hline 
\multicolumn{2}{|c|}{ Background cross section} & 
\multicolumn{3}{|c|}{30.6}    \\ \hline 
\end{tabular}
\end{center}
\caption{ Signal and background cross sections (in {\it fb}) for the 
$2 \rightarrow 4$ process $e^+ e^- \rightarrow \nu_e \bar \nu_e b \bar b$.
The resonant background estimate given in section 2 
(narrow width approximation) can be reproduced by $2 \rightarrow 4$ 
calculation with the kimematical cut of 5 GeV around $M(b \bar b)=m_Z$}. 
\end{table}

The main background to Higgsstrahlung mechanism $e^+ e^- 
\rightarrow Z^* H^*$ is given by the processes $e^+ e^- \rightarrow Z^* Z^*$,   
$e^+ e^- \rightarrow Z^* \gamma^*$ (see second row of diagrams
in Fig.1). 
Contrary to this situation the fusion mechanism
$e^+ e^- \rightarrow \nu_e \bar \nu_e H$ under the $2m_Z$ threshold is 
practically free from the background coming from $e^+ e^- \rightarrow \nu_e
\bar \nu_e Z^*$ reaction (see third row of diagrams in Fig.1). This 
observation is especially important at
the LEP II energy $\sqrt{s}=$ 175 GeV planned for the first stage of
collider operation.
 For instance, at $\sqrt{s}=$175 GeV and
$m_H$=90 GeV (5 GeV down the $m_H+m_Z$ threshold) in
the narrow width approximation we have for the signal
$$
\sigma_{tot}(e^+ e^- \rightarrow \nu_e \bar \nu_e H)*Br(H \rightarrow
b \bar b) = 7.9 \; fb
$$
while a rough estimate for the resonant background (5 GeV down the $2m_Z$ 
threshold) $$
\sigma_{tot}(e^+ e^- \rightarrow \nu_e \bar \nu_e Z)*Br(Z \rightarrow
b \bar b) = 0.79 \; fb
$$
In contrast to these numbers at $\sqrt{s}=$205 GeV the
cross sections are 41.4 fb for the Higgs signal and 30.38 fb for the
$Z$ background.
We performed more detailed calculations for the $2 \rightarrow 3$ body
background process $e^+ e^- \rightarrow \nu_e \bar \nu_e Z$ (9 Feynman 
diagrams). The contributions of two resonant graphs, the remaining 7 graphs
as well as the negative interference between these two subsets are shown
in Fig.2. If we go several GeV down the $2m_Z$ threshold the contribution
of resonant graphs decreases approximately 10 times leaving the fusion
mechanism of Higgs boson production practically background free. Signal
and background cross sections are shown in Fig. 3. 

Exact numbers (calculated using the $2 \rightarrow 4$ body matrix element 
integrated over the four particle phase space) for the two signal mechanisms 
and interference between them as 
well as for the main (resonant) background graphs at the energies 175 GeV 
and 205 GeV
are presented in Table 2. At the energy 205 GeV planned for LEP II 
upgrage (15 GeV above the $2m_Z$ threshold)
the contribution of main background graphs to the fusion mechanism increases
about 10 times. 

\section{Simulation of the signal in the detector}

We would like to consider in more details the practical
possibility of signal detection at LEP II taking into account b-quark
fragmentation and more or less realistic detector properties.

For this purpose we used the opportunity to switch on the external processes 
in the PYTHIA 5.7/JETSET 7.4 package \cite{PYTHIA}. For each event generated
by CompHEP on the partonic level six four-momenta of initial and final
particles and the total cross section were transferred to PYTHIA/JETSET 
as an input parameters. Fragmentation of b-jets, detector simulation
and jet separation were done by means of JETSET. Independent
fragmentation model (with the JETSET default parameters) was used.
Detector simulation was performed by means of standard LUCELL
subroutine (contained in JETSET).

All space available for particle detection was divided into
the cells of hadronic calorimeter. We used 64$\times$80 cells
($\varphi_0 \times \eta$) with $\eta=-ln(tg( \vartheta/2))$ and
$-4 \leq \eta \leq 4$. 
We introduced calorimeter resolution and the energy smearing in 
the detector cell. As usual the latter was defined corresponding
to gaussian distribution with the standard deviation 0.5*$\sqrt{E_{T\; 
cell}}$ with the cutoff $0 \leq E_{T\; smeared} \leq 2*E_{T\; true}$
Detector granularity was chosen to be 0.1$\times$0.1. 
The energy registered in the detector cell can be expressed as 
$$
(p_x,p_y,p_z,E,m)_{cell}=
                E_{T\;cell}*(cos\varphi,sin\varphi,sinh\eta,cosh\eta,p^2/E_{T\;cell})
$$
At the next step we separated the jets from b-quarks. All detector cells
with the energy greater than $E_{T\; cell\; min} >5$ GeV were considered
as jet initializer cells. The energy of the cells near the initializer
was summed if the distance in the $\varphi, \eta$ parameter space
$\Delta R=\sqrt{\Delta \varphi^2+ \Delta \eta^2}$ was less than $\Delta R
=$0.5 and considered as the hadronic cluster energy. If the cluster
energy was greater than $E_{T \; min}=15$ GeV the cluster was identified 
as a jet.

We represent the invariant mass distributions of $b \bar b$ at the
partonic level and the jet-jet invariant mass distributions after 
fragmentation of b-quarks and detector simulation in Fig. 4-9. At the
energy $\sqrt{s}=$ 175 GeV (Fig. 4-6) practically background free Higgs
boson peak is observed if $m_H+m_Z <$ 175 GeV . Even in the case of Higgs 
and Z-boson peaks overlap $m_H=m_Z$ which is the most complicated situation 
for the signal separation in Higgsstrahlung mechanism \cite{Brown} the 
background is practically absent . At the energy $\sqrt{s}=$ 
205 GeV (Fig. 7-9) the resonant background peak and the Higgs peak are 
observed. 

\section{Conclusion}

Our analysis shows that the cross section of Higgs boson production
under the threshold $m_H+m_Z$ in the $e^+ e^- \rightarrow \nu \bar \nu b 
\bar b$ channel receives the very important contribution from the fusion 
mechanism additionally enhanced by the positive Higgsstrahlung-fusion 
interference. The cross section is rather small and the Higgs peak is 
strongly smeared by b-jet fragmentation process and the effects of 
limited detector resolution. Probably it would be hardly possible to 
reconstruct precisely the mass of Higgs resonance using the invariant 
mass distribution of two b-jets. However it seems realistic to
observe some strong indication to the Higgs boson production at 
LEP II energies. The energy point $\sqrt{s}=$ 175 GeV in especially
interesting for observation of the Higgs boson under the threshold. 
Detailed study of the background shows that at 
this energy (below the $2m_Z$ threshold) the fusion mechanism of Higgs
boson production is almost background free.
  
\begin{center}
{\bf Acknowledgements}
\end{center}
The authors express their gratitude to the
Centre of Theoretical Physics, Seoul National University where this work
was completed. The research was partially supported by INTAS grant 93-1180.

\newpage

\unitlength=0.8pt

\begin{figure}[tbp]
\input{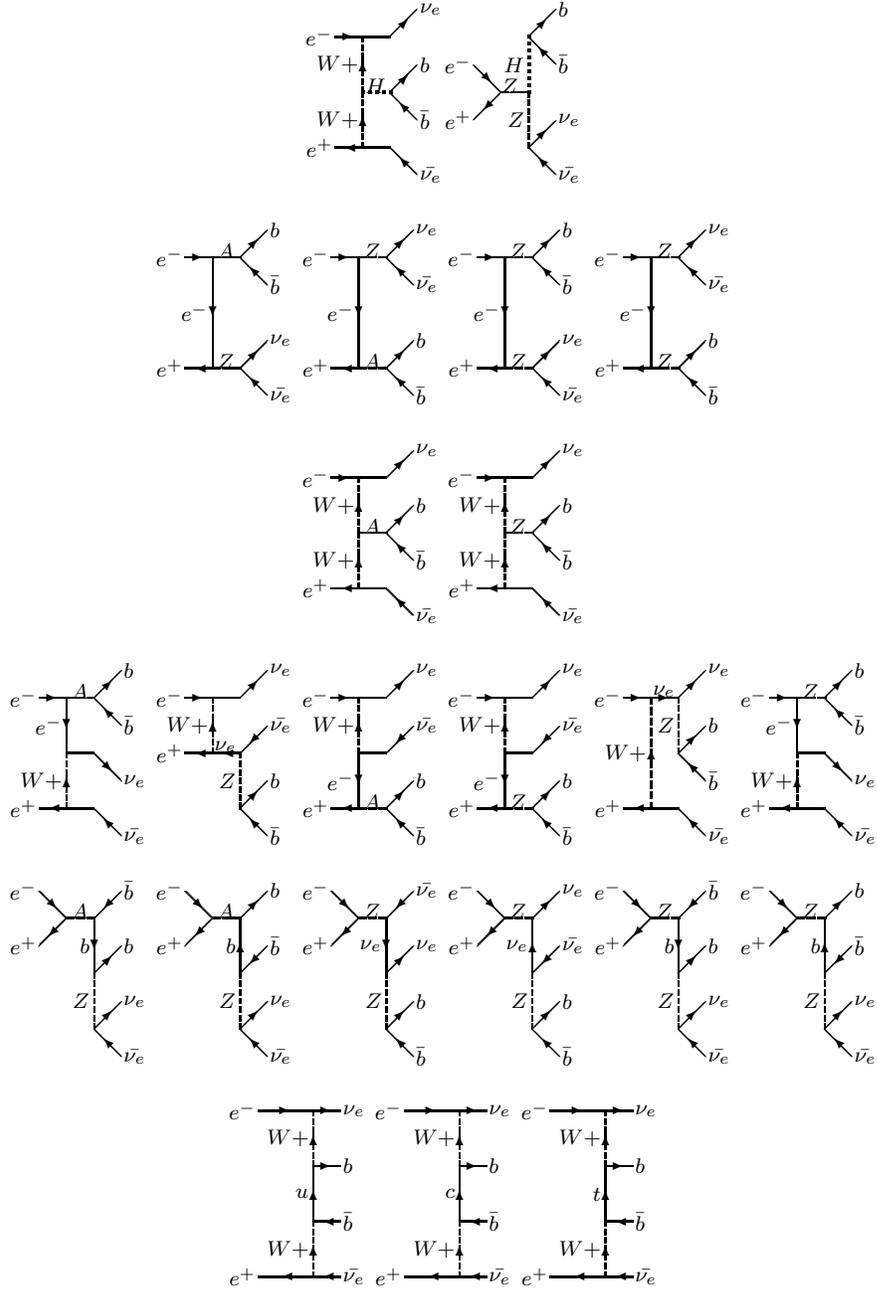}
\caption{Complete set of diagrams for the process
$e^+ e^- \rightarrow \nu_e \bar \nu_e b \bar b$}
\end{figure}

\newpage

\unitlength=10mm

\begin{figure}[tbp]
\begin{picture}(10,10)
\put(0,-3){\epsfxsize=12.5cm \leavevmode \epsfbox{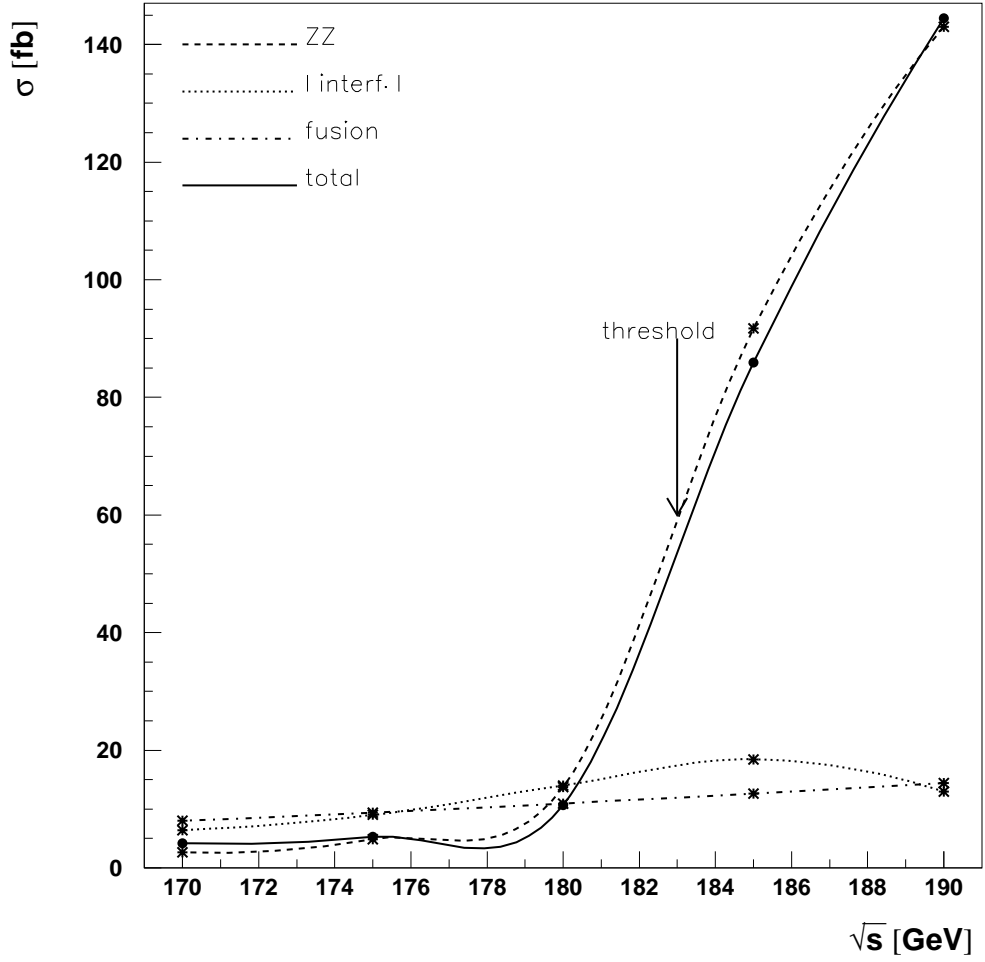} }
\end{picture}
\caption{Total cross sections for the two subsets of diagrams 
(s-chanel resonant diagrams as the first subset and all the rest as the 
second) and the interference between them in the process $e^+ e^- \rightarrow
\nu_e \bar \nu_e Z$. The absolute value of negative interference term is 
shown.} 
\end{figure}

\newpage

\begin{figure}[tbp]
\begin{picture}(10,10)
\put(0,-3){\epsfxsize=12.5cm \leavevmode \epsfbox{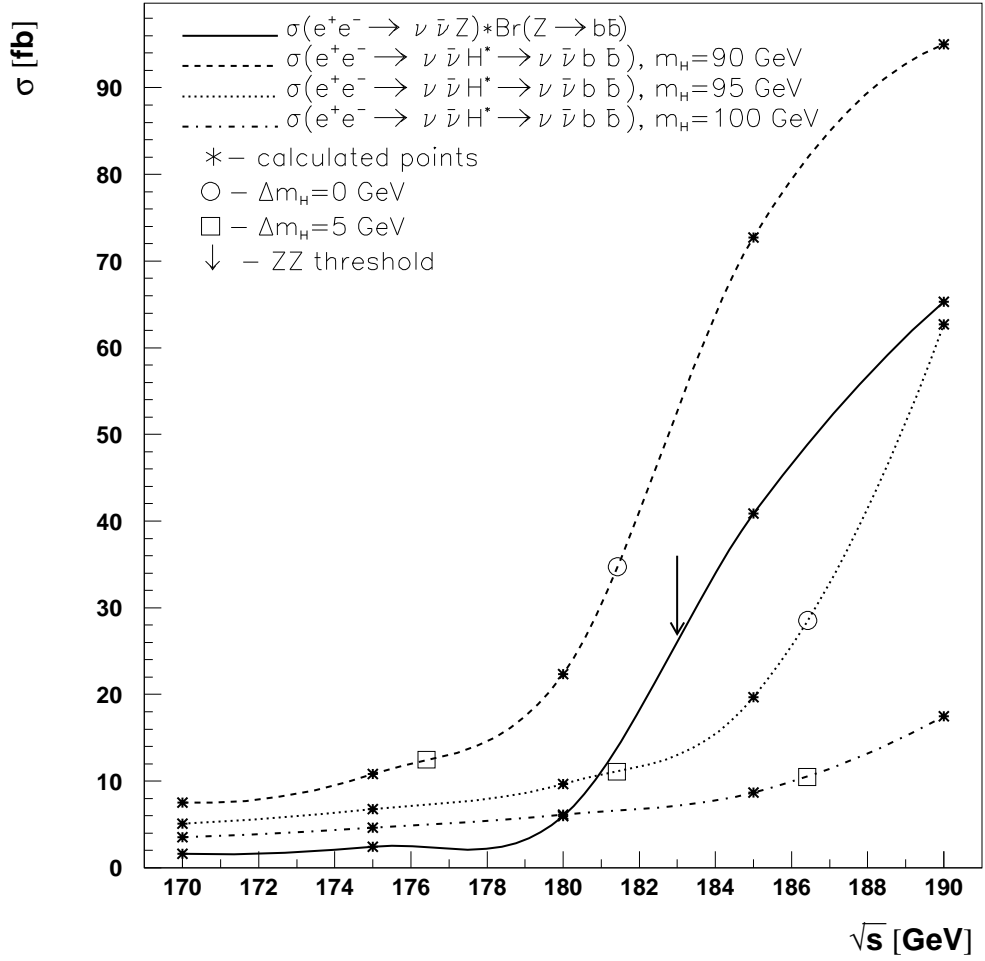} }
\end{picture}
\caption{Background total cross section from Fig.2 together with
the signal cross section $e^+ e^- \rightarrow \nu_e \bar \nu_e b \bar b$.
The distance from the $HZ$ threshold $\Delta m = -{\protect \sqrt{s}}+m_H+m_Z$
in the cases $m_H=$ 90, 95, 100 GeV is marked on the corresponding cross
section curves by the circles ($\Delta m =$0) and squares ($\Delta m =$ 5 
GeV).} 
\end{figure}

\newpage

\begin{figure}[tbp]
\begin{picture}(10,10)
\put(0,-2){\epsfxsize=12.5cm \leavevmode \epsfbox{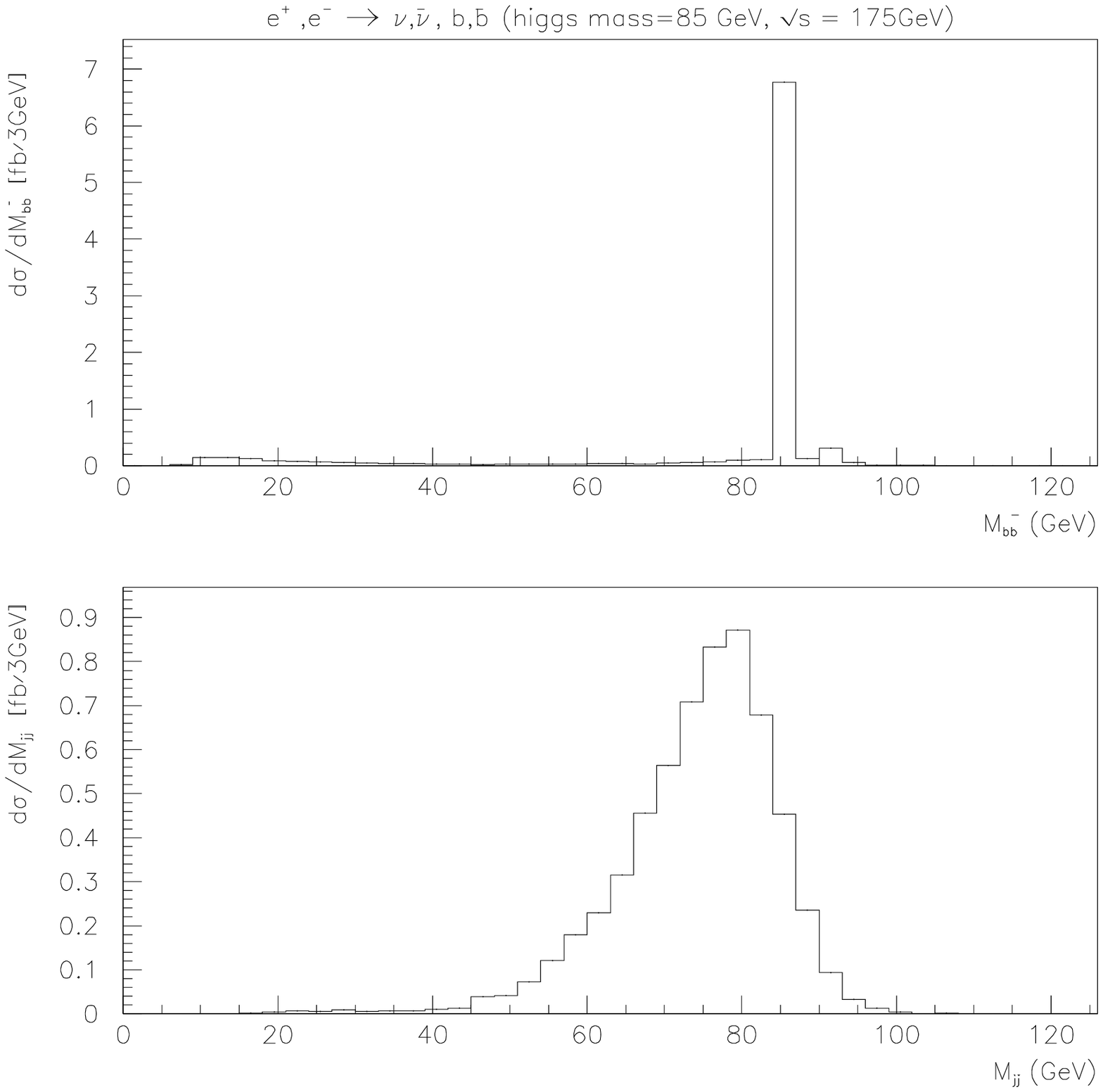} }
\end{picture}
\caption{ }
\end{figure}

\newpage

\begin{figure}[tbp]
\begin{picture}(10,10)
\put(0,-2){\epsfxsize=12.5cm \leavevmode \epsfbox{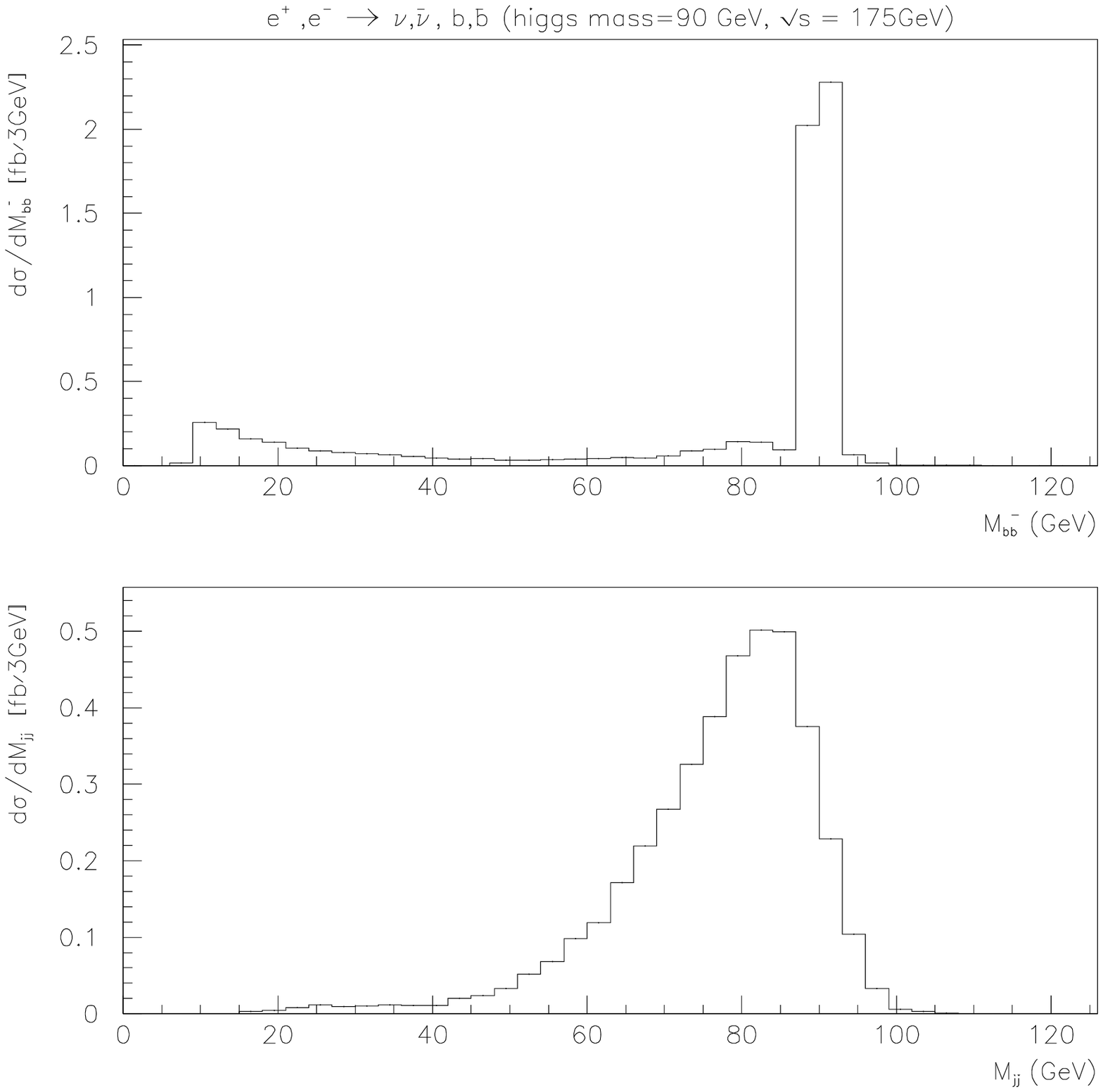} }
\end{picture}
\caption{ }
\end{figure}

\newpage

\begin{figure}[tbp]
\begin{picture}(10,10)
\put(0,-2){\epsfxsize=12.5cm \leavevmode \epsfbox{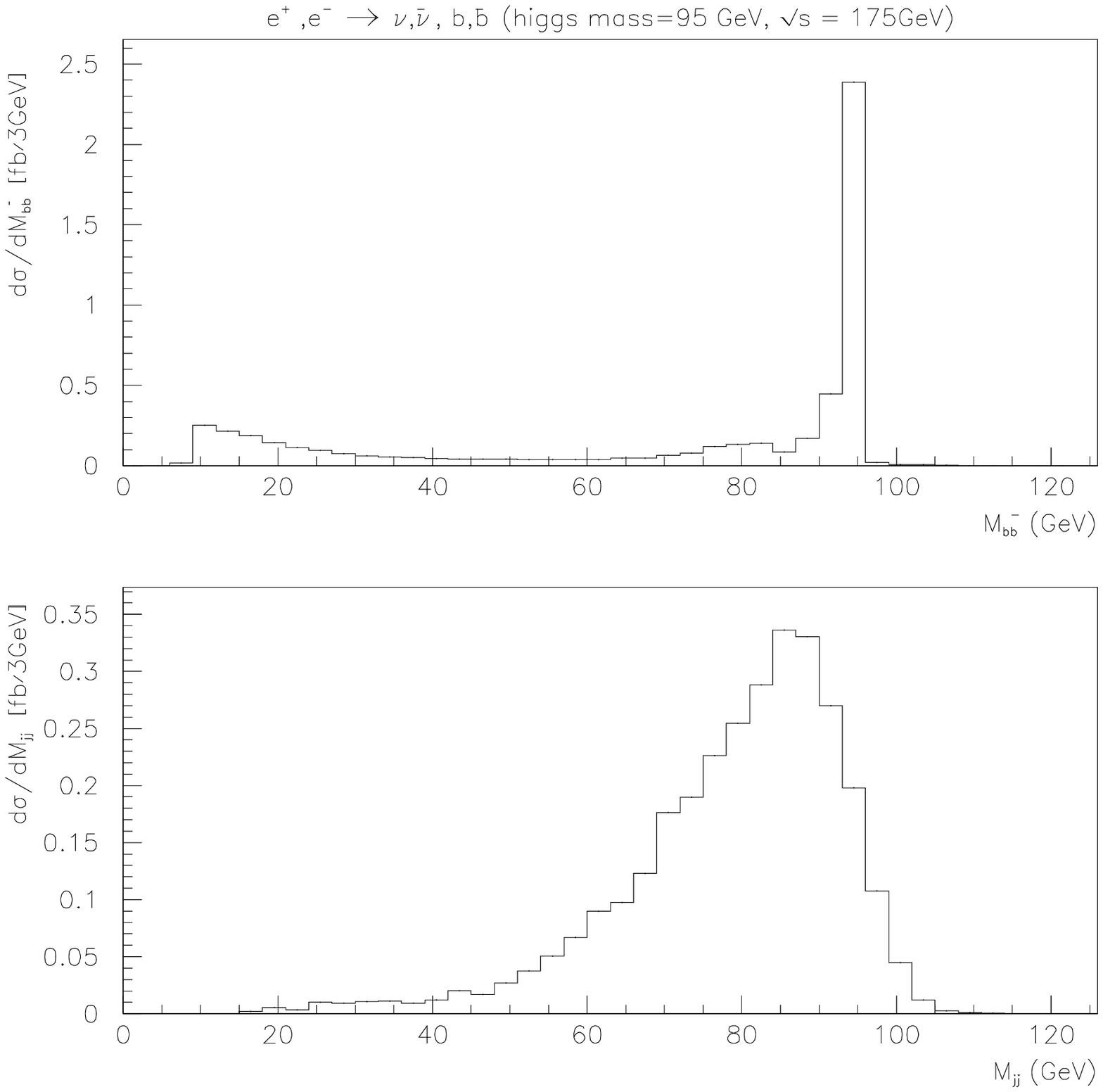} }
\end{picture}
\caption{ }
\end{figure}

\newpage

\begin{figure}[tbp]
\begin{picture}(10,10)
\put(0,-2){\epsfxsize=12.5cm \leavevmode \epsfbox{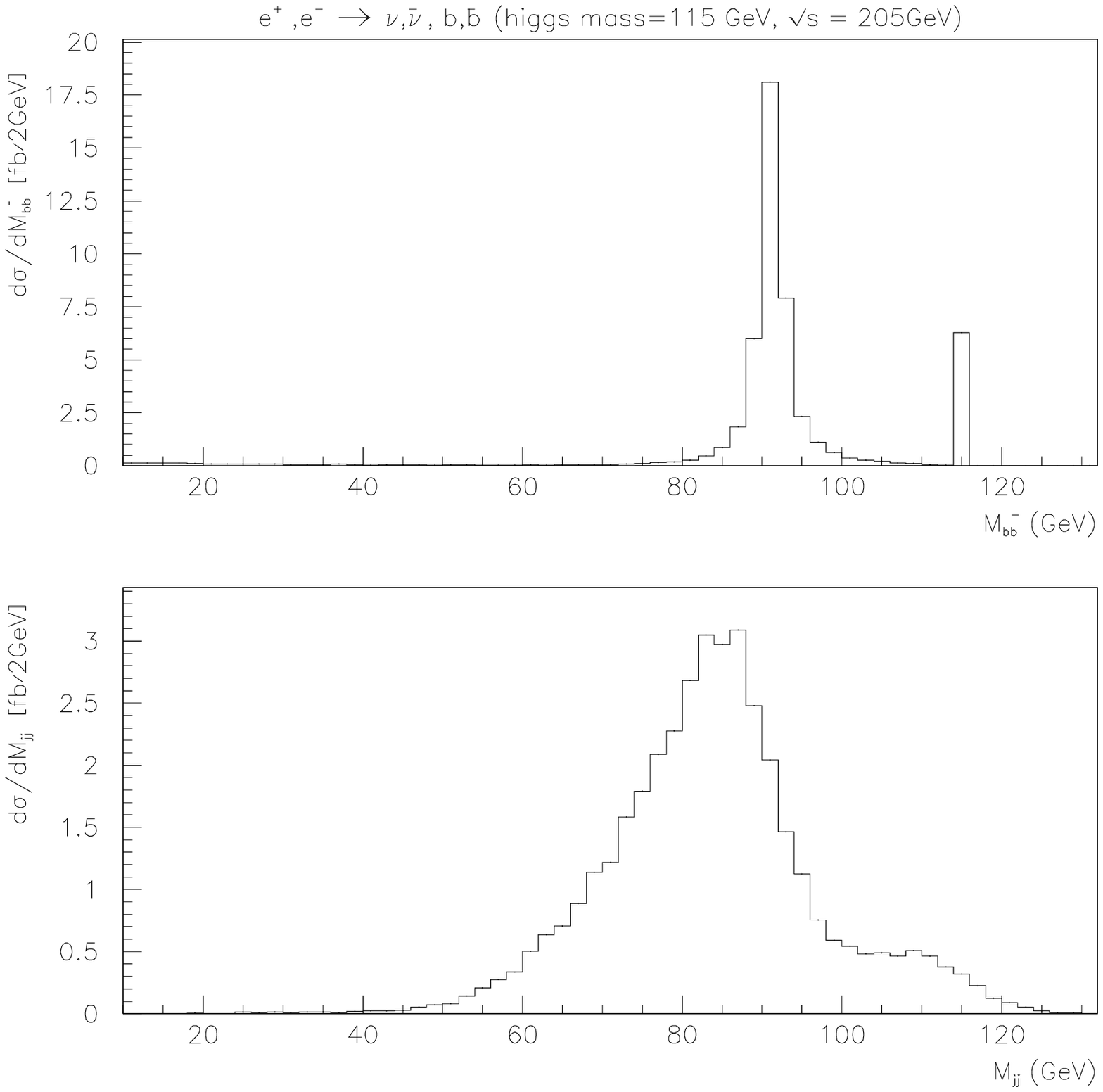} }
\end{picture}
\caption{ }
\end{figure}

\newpage

\begin{figure}[tbp]
\begin{picture}(10,10)
\put(0,-2){\epsfxsize=12.5cm \leavevmode \epsfbox{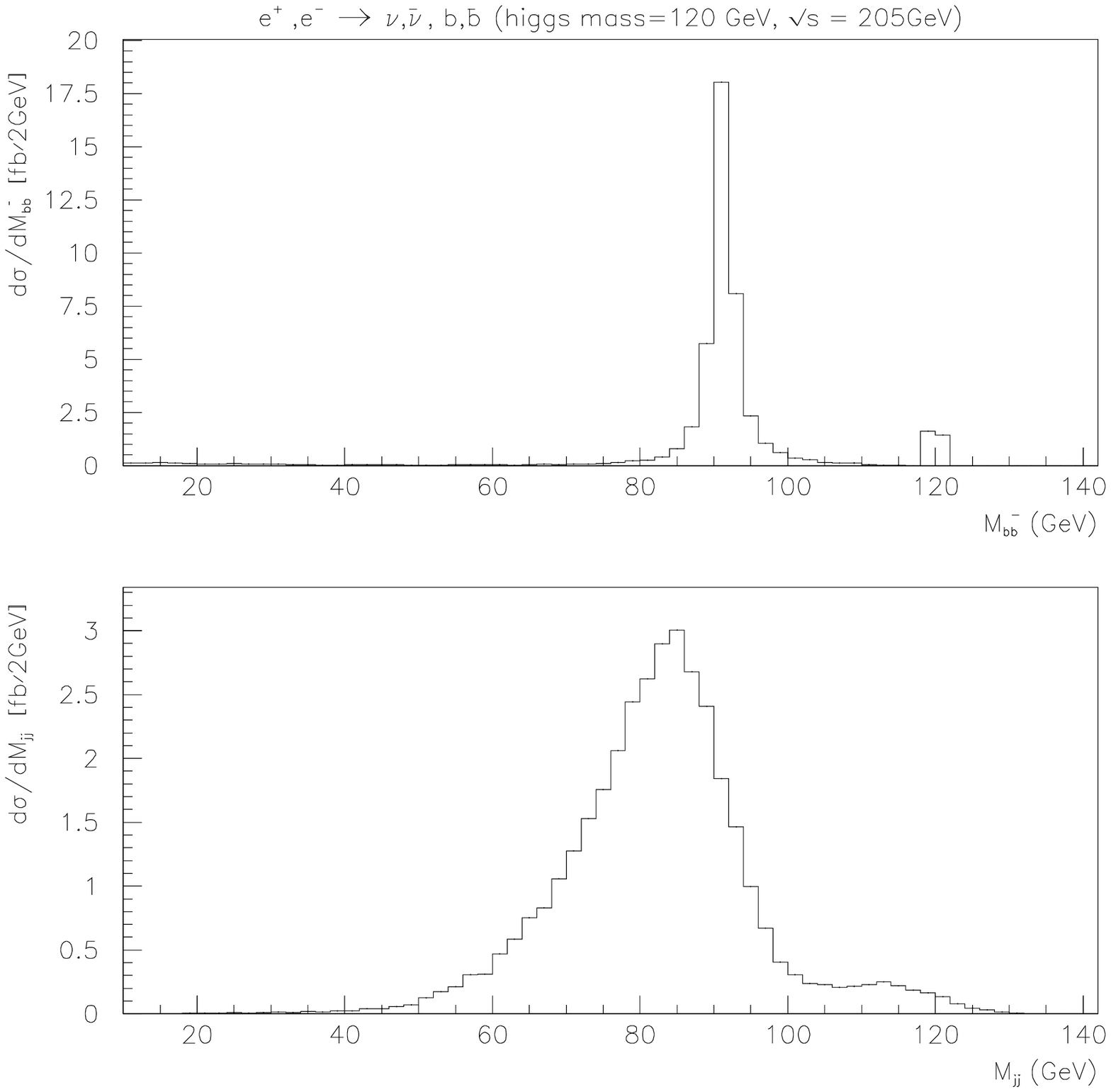} }
\end{picture}
\caption{ }
\end{figure}

\newpage
 
\begin{figure}[tbp]
\begin{picture}(10,10)
\put(0,-2){\epsfxsize=12.5cm \leavevmode \epsfbox{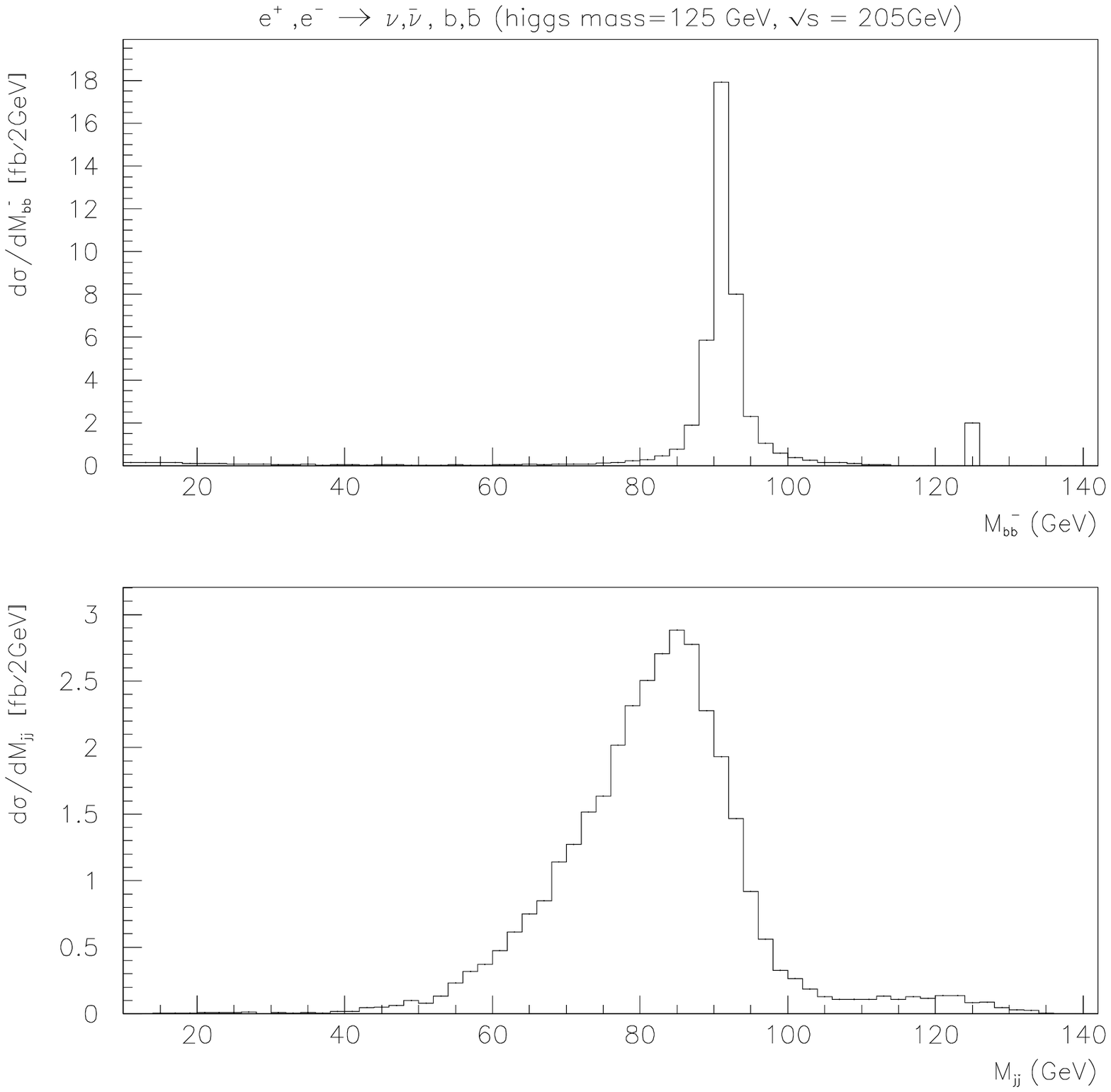} }
\end{picture}
\caption{ }
\end{figure}


\newpage
 
\begin{thebibliography}{99} 

\bibitem{BSSS1} 
E.Boos, M.Sachwitz, H.J.Schreiber, S.Shichanin, Z.Phys. C61 (1994) 675 \\
G.Montagna, O.Nicrosini, F.Piccinini, Phys.Lett. B348 (1995) 496

\bibitem{BSSS2} 
M.Dubinin, V.Edneral, Y.Kurihara, Y.Shimizu, Phys.Lett. B329 (1994) 379 \\
E.Boos, M.Sachwitz, H.J.Schreiber, S.Shichanin, Int.J.Mod.Phys. A10 
(1995) 2067 

\bibitem{BSSS3} 
E.Boos, M.Sachwitz, H.J.Schreiber, S.Shichanin, Z.Phys. 
C64 (1994) 391 

\bibitem{brems}
J.D.Bjorken, in: Proc.of Summer Institute on Particle Physics, ed.by
M.Zipf, Stanford, 1976  \\
J.Ellis, M.K.Gaillard, D.V.Nanopoulos, Nucl.Phys. B106 (1976) 292 \\
B.L.Ioffe, V.A.Khoze, Phys.Elem.Part.At.Nucl.(USSR) 9 (1978) 118

\bibitem{fusion}
D.R.T.Jones, S.T.Petcov, Phys.Lett. B84 (1979) 440 \\
G.Altarelli, B.Mele, F.Pitolli, Nucl.Phys., B287 (1987) 205

\bibitem{BD} 
E.Boos, M.Dubinin, Phys.Lett. B308 (1993) 147 

\bibitem{KKZ} 
W.Kilian, M.Kramer, P.Zerwas, DESY preprint 95-216, 1995
 (hep-ph/9512355)

\bibitem{Bardin}
D.Bardin, A.Leike, T.Riemann, Nucl.Phys. B, Proc.Suppl. 37B (1994) 274 \\
D.Bardin, A.Leike, T.Riemann, Phys.Lett. B344 (1995) 383 \\
D.Bardin, A.Leike, T.Riemann, Phys.Lett. B353 (1995) 513 

\bibitem{general}
F.A.Berends, R.Pittau, R.Kleiss, Nucl.Phys. B424 (1994) 308 \\
T.Ishikawa, T.Kaneko, K.Kato, S.Kawabata, Y.Shimizu, H.Tanaka,
 KEK report 92-19, 1993

\bibitem{CompHEP}
E.Boos et al., Proc.of the XXVIth Recontre de Moriond, ed.by J.Tran Thanh
Van, Editions Frontiers, 1991, p.501 \\
E.Boos et al., Proc.of the Int.Conf. on Computing in High Energy
Physics, ed.by Y.Watase, F.Abe, Universal Academy Press, Tokyo,
1991, p.391 \\
E.Boos et al., SNU CTP preprint 94-116, Seoul, 1994 (hep-ph/9503280)

\bibitem{width}
A.Aeppli, F.Cuypers, Geert van Oldenborgh, Phys.Lett., B314 (1993) 413 \\
E.Argyres, W.Beenakker, Geert van Oldenborgh, A.Denner, \\
S.Dittmaier, J.Hoogland, R.Kleiss, C.Papadopoulos, G.Passarino, INLO-PUB-8/95 
(hep-ph/9507216)

\bibitem{PYTHIA}
T.Sjoestrand, M.Bengtsson, Comp.Phys.Comm., 43 (1987) 367 \\
H.-U. Bengtsson, T.Sjoestrand, Comp.Phys.Comm., 46 (1987) 43
 
\bibitem{Brown}
Z.Kunszt, W.Stirling, Phys.Lett., B242 (1990) 507 \\
N.Brown, Z.Phys., C49 (1991) 657

\end{thebibliography}
\end{document}